# Analysis of Generalized Impact Factor and Indices of Journals

Ash Mohammad Abbas
Department of Computer Engineering
Aligarh Muslim University
Aligarh - 202002, India
Email: am.abbas.ce@amu.ac.in

*Abstract*—Analyzing the relationships among the parameters for quantifying the quality of research published in journals is a challenging task. In this paper, we analyze the relationships between impact factor, $h$-index, and $g$-index of a journal. To keep our analysis simple and easy to understand, we consider a generalized version of the impact factor where there is no time window. In the absence of the time window, the impact factor converges to the number of citations received per paper. This is not only justified for the impact factor, it simplifies the analysis of $h$-index and $g$-index as well because addition of a time window in the form of years complicates the computation of indices too. We derive the expressions for the relationships among impact factor, $h$ index, and $g$-index and validate them using a given set of publication-citation data.

## I. INTRODUCTION

Sometimes, one needs to rank the journals where the outcomes of the research carried out by authors working in a particular field of research are published. The ranking of the journals may vary depending upon which parameter is selected for ranking. Generally, the journals are ranked based on the parameters that are derived from the citations of the papers published in the journals. One such parameter is the impact factor which tells about the number of citations divided by the number of papers published in a constant number of the preceding years. Another parameter is the $h$-index that tells about the how many papers published in the journal possess at least the same number of citations as that of the number of papers. Yet, another parameter is the $g$-index, which is the largest number so that the summation of the citations is at least the square of the number, and this applies only when papers are arranged in the decreasing number of their citations.

Although, these parameters seems to be different, however, they might be related in some sense. There is a need to investigate the relationships among these parameters so that given the value(s) of some parameter(s), one can determine the other ranking parameter. Alternatively, from a set of values of one ranking parameter, one is able to predict the values of other ranking parameter. Sometimes, the analysis of the relationships among different ranking parameters enables one to get clues why the rankings of the journals differ by changing the parameter used for ranking.

Many researchers have tried to investigate the relationships among different ranking parameters and for journals in different domains. A comparative analysis between impact factor and $h$-index for pharmacology and psychiatry journals is carried out in [2]. Therein, a hypothesis for modelling the relationship between $h$-index and impact factor of a journal is discussed assuming that citation rate of a paper is a random variable and follows the Pareto distribution.

The $g$-index is proposed in [9]. An analysis of $g$-index is described in [3]. A relationship between $h$-index and $g$-index is discussed using Lotka's model, $f(j) = \frac{C}{j^\alpha}$, where, $j \geq 1, C > 0, \alpha > 2$. In [4], an analysis of the relationship between impact factor and uncitedness is carried out assuming that the publication-citation relationship follows Lotka's model. A relationship between impact factor, $h$-index, and $g$-index using power law model is described in [5]. A relationship between $h$-index, $g$-index, and $e$-index is described in [7], where indices are assumed to be modelled as continuous functions.

In this paper, we analyze the relationships among the impact factor, $h$-index, and $g$-index of a given journal. We assume that the impact factor of a given journal is average number of citations of the paper published in the journal. This assumption seems to be realistic because the impact factor of a journal, in the long term, is nothing but the average number of citations per paper. The same assumption is used in [5], where a relationship between the impact factor and $h$-index is described using Lotka's power law model. Our work is different from [5] in the sense that we do not use a specific model, such as Lotka's model, to derive the relationship among the impact factor and the indices. Moreover, our work is different from [7] in the sense that in [7], the relationships among $h$-index, $g$-index, and $e$-index [6], are analyzed and not the impact factor. However, we analyze the relationships among the impact factor, $h$-index, and $g$-index. Further, as opposed to [7], where indices are assumed to be represented by continuous functions and the analysis is centered around the $e$-index; we use the original definitions of indices which are discrete in nature, and our analysis is focused around the impact factor of a journal. In other words, we start from the definitions of indices and the impact factor and derive relationships among them without assuming that either of them follows a specific distribution or is represented by a continuous or smooth function.

Although, we focus on the analysis of the relationships among the impact factor, $h$-index, and $g$-index of a journal. However, our approach is not merely confined to journals.

For example, the concept of a long term impact factor can be applied to individuals, research groups, departments, universities, countries, and continents; and also to venues such as conferences, workshops, proceedings, books, publishers, etc. The only thing is that how we view the groups and venues in terms of their research outputs. Alternatively, as for the $h$-index and $g$-index, which can be applied to groups or venues, the long term impact factor can also be applied to groups or venues. For example, the average number of citations per paper published by an individual researcher is an impact factor of the researcher, the average number of citations per papers published by a research group is the impact factor of the research group; and continuing in this manner, the average number of citations per papers published by a university is the impact factor of the university, and so on.

The rest of the paper is organized as follows. In Section II, we present an overview of the impact factor and indices. Section III contains the analysis of the relationships among the impact factor and indices. In Section IV, we present results and discussion. The last section is for conclusions.

## II. AN OVERVIEW OF INDICES AND IMPACT FACTOR

In this section, we present an overview of the indices and impact factor so as to prepare a background for understanding the concepts presented in this paper.

### A. H-Index

Suppose the papers are arranged in descending order of the number of citations. Let $c_i$ be the number of citations of a paper numbered $i$. The $h$-index [8], when papers are arranged in descending number of their citations, can be defined as follows.

$$h = \max(i) : c_i \geq i. \quad (1)$$

By definition, $h$-index is the largest number, $h$, such that the papers arranged in their decreasing order of citations have at least $h$ number of citations. Therefore, for $i = h$, $c_i = h$, and for $i = h + 1$, $c_i < h + 1$. Assume that the papers are arranged in the descending order of their citations. If one plots the number of citations as a function of the paper number, the line joining the points $(i, c_i) = (0, 0), (h, h)$ makes an angle of $45°$ from the $x$ and $y$ axes, and crosses the citation curve so drawn at the point $(h, h)$.

### B. G-Index

According to the definition of $g$-index, if the papers are arranged in the descending order of their number of citations, $g$ is the largest number such that the summation of the number of citations is at least $g^2$. In other words, when papers are arranged in descending order of their citations, $g$-index can be defined as follows.

$$g = \max(i) : \sum_i c_i \geq i^2. \quad (2)$$

*Example 1:* Let the number of citations, $c_i$, of paper numbered, $i$, be as shown in Table I, where the papers are arranged in the descending order of their number of citations. Compute the $h$-index and $g$-index for these set of values.

TABLE I
EXAMPLE 1: PAPER NUMBER, $i$, NUMBER OF CITATIONS, $c_i$, SUMMATION OF CITATIONS, $\sum_i c_i$, AND $i^2$ (FICTITIOUS EXAMPLE).

| $i$ | $c_i$ | $\sum_i c_i$ | $i^2$ |
|---|---|---|---|
| 1 | 30 | 30 | 1 |
| 2 | 24 | 54 | 4 |
| 3 | 18 | 72 | 9 |
| 4 | 14 | 86 | 16 |
| 5 | 12 | 98 | 25 |
| 6 | 11 | 109 | 36 |
| 7 | 10 | 119 | 49 |
| 8 | 9 | 128 | 64 |
| 9 | 8 | 136 | 81 |
| 10 | 7 | 143 | 100 |
| 11 | 6 | 149 | 121 |
| 12 | 5 | 154 | 144 |
| 13 | 4 | 158 | 169 |
| 14 | 3 | 161 | 196 |
| 15 | 2 | 163 | 225 |
| 16 | 1 | 165 | 256 |

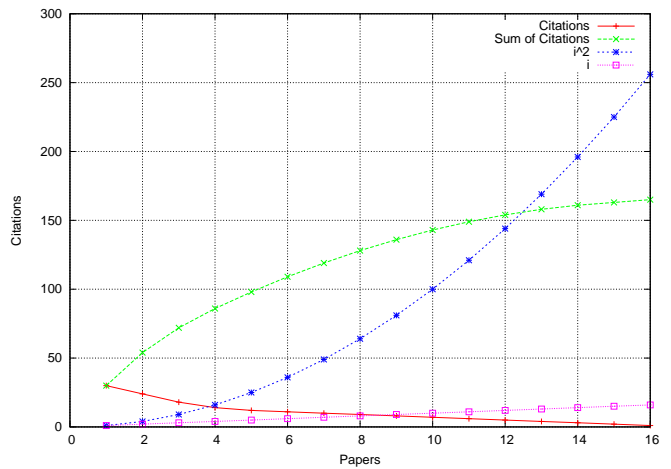

Fig. 1. The $g$-index for a given set of citations of Example 1. If the papers are arranged in the descending order of their citations, the $g$-index is the largest integer, $i$, where $\sum_i c_i \geq i^2$.

The $h$-index is 8 because 8 papers have at least 8 citations. Alternatively, the curve showing the number of citations, $c_i$, crosses the line for number of papers equal to $i$ after $i = 8$. The $h$-index, which is the largest integer $i$ such that $c_i \geq i$ (as defined by (1)), is 8. The $g$-index is 12, because $i = 12$ is the largest number for which summations of the number of citations of the first $i$ papers is greater than or equal to $i^2$. This is also illustrated in Figure 1. Note that $g$-index is the largest number $i$ such that $\sum_i c_i \geq i^2$. The $g$-index in Figure 1 is the integer number, $i$, which is just before the point where the $\sum_i c_i$ curve ceases to exceed the the curve $i^2$, (or $i^2$ curve starts dominating the $\sum_i c_i$ curve). In this case, the value of $g$-index is 12.

In what follows, we describe a notion of the generalized impact factor.

### C. Impact Factor

Generally, the impact factor of a journal is defined using a time window. For example, an impact factor may be computed for a time window of either five years or two years, and are termed as *five year impact factor* or *two year impact factor*, respectively. To understand how the impact factor is actually computed, assume the impact factor is computed on two years basis. The expression for the two year impact factor can be described as follows.

*Definition 1 (Impact Factor on Two Year Basis):* Let the number of papers published by the journal in year $y_1$ be $P_{y_1}$, and in year $y_2$ be $P_{y_2}$. Let the number of citations in the year $y_3$, which is successor of the year $y_2$, for the papers published in years $y_1$ and $y_2$ be $C_{y_3}$. Then, the impact factor of the journal for the year $y_3$ is as follows.

$$(IF)_{y_3} = \frac{C_{y_3}}{P_{y_1} + P_{y_2}}. \quad (3)$$

We now provide a general definition of the impact factor with a time window constraint.

*Definition 2 (Impact Factor with a Time Window Constraint):* Let $W$ be the time window for computing the impact factor, and let $y_b$ be the starting (or the base) year for computing the impact factor. Then, the impact factor of a journal, in general, can be defined as follows.

$$(IF)_{y_{\{b+W\}}} = \frac{C_{y_{\{b+W\}}}}{\sum_{i=0}^{W-1} P_{y_{\{b+i\}}}}. \quad (4)$$

We say that this definition of impact factor is *general* in the sense that it is able to incorporate any time window. For example, if $W = 2$, the impact factor is on two year basis; and $W = 5$ makes the impact factor on five year basis. Moreover, one is not confined to only these two values, as one can choose any other value of $W$. In the following, we explain how one can express an impact factor on five year basis through an example.

*Example 2 (Impact Factor on Five Year Basis):* Let the starting year, $y_b$, be 2001. Assume that we consider a five year impact factor. We wish to write an expression for the five year impact factor.

From the year 2000 to year 2005, there are 5 years (including year 2001 and year 2005). Using (4) the impact factor of year 2006 is given as follows.

$$\begin{aligned}(IF)_{2006} &= \frac{C_{\{2001+5\}}}{\sum_{i=0}^{5-1} P_{\{2001+i\}}} \\ &= \frac{C_{2006}}{P_{2001} + P_{2002} + P_{2003} + P_{2004} + P_{2005}}\end{aligned} \quad (5)$$

Similarly, using (4) one can write an expression for an impact factor on the basis of a given number of years.

As we mentioned earlier, our goal in this paper is to relate the impact factor of a journal with the indices. Specifically, we wish to find out a relationship between the impact factor, $h$-index, and $g$-index. To relate them, either $h$-index and $g$-index should be defined taking the same time window as for that of the impact factor, or the time window should be eliminated from the impact factor so that all these parameters become coherent. Otherwise, a comparison between them (without making them coherent) may result into a comparison of *apples* and *potatoes*. Note that introducing the *time window* in the definition of $h$-index and $g$-index may not make these parameters completely coherent and may result in unnecessarily complications in the definitions of $h$-index as well as the $g$-index. A simpler and better alternative seems to be the removal of time window from the definition of impact factor. We can replace the time window in the impact factor with the phrase "till now". That is, we can say that the impact factor is the number of citations received by the journal "till now" divided by the number of papers published in the journal "till now". Note that in the definitions of the original $h$-index [8] and that of the $g$-index [9], the phrase "till now" is hidden for a journal, unless one considers variations of these indices that incorporate the time windows for the indices of journals as well. Therefore, we can say that the definition of an impact factor without a time window constraint is coherent with the definitions of $h$-index as well as $g$-index both without any time window constraints. We now define an impact factor that we call a *generalized impact factor* as follows.

*Definition 3 (Impact Factor without a Time Window Constraint):* Let the total number of papers published in the journal be $P$ and the total number of citations received by the journal be $C$. The *generalized impact factor* or an impact factor without a time window constraint is as follows.

$$I_f = \frac{C}{P}. \quad (6)$$

The *generalized impact factor*, $I_f$, resembles with the average number of citations of the journal per published paper, and that is in accordance with the definition of the impact factor. In other words, if the time window constraint is removed, the impact factor turns out to be the average number of citations per published paper. We wish to point out that we are not the first ones who adopt a definition of the impact factor without any time window constraints, there are other researchers such as [5] who have also taken into account the similar kind of definition (i.e. without any time window) of the impact factor, and who agree that there is no harm in taking this type of definition for the purpose relating indices and the impact factor. The reason is that the definition without a time window constraint puts aside the complications of redefining indices from a window-less scenario to a windowed scenario so as to make them coherent with the impact factor with a time window constraint.

Now, coming back to the example discussed above (Example 1) for computing the $g$ index, the *generalized impact factor*, $I_f$, is equal to $\frac{165}{16} = 10.3125$.

## III. IMPACT FACTOR AND $h$-INDEX

Let $P$ be the number of papers published in a journal, and let $c_i$ be the number of citations of $i$th paper. Then, without considering a time window, the *impact factor* ($I_f$) of a journal can be expressed as follows.

$$I_f = \frac{\sum_i c_i}{P}. \quad (7)$$

According to the definition of $h$-index, $h$ papers have at least $h$ number of citations, therefore, $h^2$ citations are taken into account by the $h$-index. In other words, if a journal has an $h$-index, $h$, then $h^2$ of the citations are taken care of by the $h$-index. The rest of the citations are not taken into account by the $h$-index. We can write the total citations as follows.

$$\sum_{i=1}^{P} c_i = h^2 + \sum_{i=1}^{h}(c_i - h) + \sum_{i=h+1}^{P} c_i. \qquad (8)$$

Using (7), we can write (8) as follows.

$$I_f \times P = h^2 + \sum_{i=1}^{h}(c_i - h) + \sum_{i=h+1}^{P} c_i. \qquad (9)$$

Or,

$$I_f = \frac{1}{P}\left[h^2 + \sum_{i=1}^{h}(c_i - h) + \sum_{i=h+1}^{P} c_i\right]. \qquad (10)$$

## IV. IMPACT FACTOR AND $g$-INDEX

Using the definition of $g$-index, which is given by (2), we have,

$$g = \max(i) : \sum_i c_i \geq i^2.$$

The above equation can be written as

$$\sum_{i=1}^{g} c_i \geq g^2. \qquad (11)$$

As we did for $h$ index, breaking the total number of citations into two parts, one ranging from 1 to $g$, and the other ranging from $g+1$ to $P$, we have,

$$\sum_{i=1}^{P} c_i = \sum_{i=1}^{g} c_i + \sum_{i=g+1}^{P} c_i. \qquad (12)$$

Combining (11) and (12), we have,

$$\sum_{i=1}^{P} c_i = g^2 + \sum_{i=g+1}^{P} c_i. \qquad (13)$$

Using (13) and (7), we have,

$$I_f \times P = g^2 + \sum_{i=g+1}^{P} c_i. \qquad (14)$$

Or,

$$I_f = \frac{1}{P}\left[g^2 + \sum_{i=g+1}^{P} c_i\right]. \qquad (15)$$

Using (10) and (15), we have,

$$\frac{1}{P}\left[h^2 + \sum_{i=1}^{h}(c_i - h) + \sum_{i=h+1}^{P} c_i\right] = \frac{1}{P}\left[g^2 + \sum_{i=g+1}^{P} c_i\right]. \qquad (16)$$

Or,

$$h^2 + \sum_{i=1}^{h}(c_i - h) + \sum_{i=h+1}^{P} c_i = g^2 + \sum_{i=g+1}^{P} c_i. \qquad (17)$$

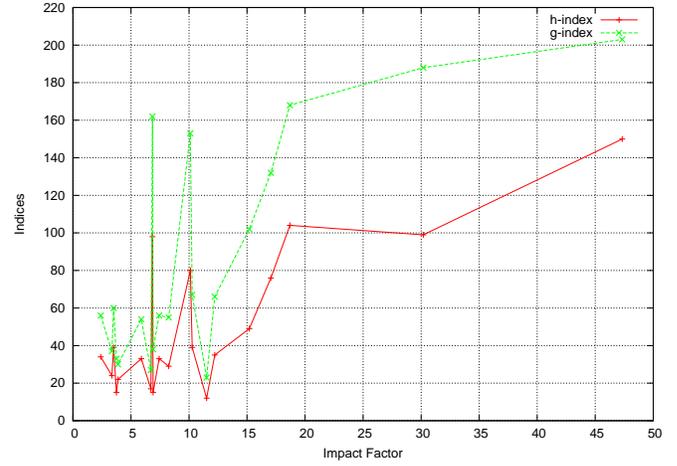

Fig. 2. The values of $h$-index and $g$-index as a function of the impact factor, where journals are arranged according to the increasing values of their impact factors.

Taking out the summations on one side, (17) can be written as

$$g^2 - h^2 = \sum_{i=1}^{h}(c_i - h) + \sum_{i=h+1}^{P} c_i - \sum_{i=g+1}^{P} c_i. \qquad (18)$$

If we assume that $g \geq h$, we find the L.H.S. of (18) is $+ve$. What about the R.H.S.? The answer to this question can be explained as follows. Consider the following difference of summations.

$$\sum_{i=h+1}^{P} c_i - \sum_{i=g+1}^{P} c_i. \qquad (19)$$

Since $g \geq h$, we have, $\sum_{i=h+1}^{P} c_i \geq \sum_{i=g+1}^{P} c_i$. This is because the gap among the citations from $h+1$ to $P$ is greater than or equal to that from $g+1$ to $P$. As a result, (19) comes out to be $+ve$, and the first term in the R.H.S. of (18) is $+ve$. Therefore, the R.H.S. of (18) is $+ve$, and this verifies the correctness of (18).

## V. RESULTS AND DISCUSSION

We computed generalized impact factor, $h$-index, and $g$-index for journals based on the citations in the Microsoft Academic Search [11] (MAS). A reason for choosing MAS is that it is freely accessible. The impact factor and indices are listed for top ranked journals in the *networks and communication* group of the Computer Science area and are shown in Table II. Journals are ranked according to the decreasing values of their impact factors.

Figure 2 shows the $h$-index and $g$-index of the journals considered in this paper (as given in Table II) as a function of the impact factor. Note that for Figure 2 journals are arranged in the increasing order of their impact factors. We observe that as the impact factor increases, the $h$-index and the $g$-index also increase, in general. In other words, a larger impact factor, in general, means larger values of $h$-index and $g$-index. For some of the journals, even though the impact factor is small, however, the values of $h$-index and $g$-index are comparatively large. A closer look on Figure 2 in conjunction with Table II reveals that it happens in case of those journals which have a large number of citations, $\sum_i c_i$, and a large number of papers published, $P$. As a result, the impact factor which is taken to be the average number of citations per paper is small, however, there is a fairly large number of papers to increase the $h$-index as well as the $g$-index. Another point to observe from Figure 2 is that the journals that possess a large value of $h$-index also possess a large value of the $g$-index. This can be understood on the basis of (18), which implies a larger value of $g$-index for a larger value of $h$-index, and vice versa.

TABLE II
THE IMPACT FACTOR, $h$-INDEX, AND $g$-INDEX OF JOURNALS IN NETWORKS AND COMMUNICATION GROUP.

| S.No. | Journal | Acronym | $\sum_i c_i$ | $P$ | $I_f$ | $h$-index | $g$-index |
|---|---|---|---|---|---|---|---|
| 1 | ACM Computer Communication Review | CCR | 85809 | 1813 | 47.330 | 150 | 203 |
| 2 | IEEE Transactions on Networking | ToN | 49148 | 1628 | 30.189 | 99 | 188 |
| 3 | IEEE Journal on Selected Areas in Communication | JSAC | 64330 | 3441 | 18.695 | 104 | 168 |
| 4 | ACM Performance Evaluation Review | PER | 27795 | 1630 | 17.052 | 76 | 132 |
| 5 | Wireless Networks | WINET | 13588 | 894 | 15.199 | 49 | 102 |
| 6 | ACM Mobile Computing & Communication Review | MCCR | 5304 | 434 | 12.221 | 35 | 66 |
| 7 | Journal of Communication & Networks | JCN | 645 | 56 | 11.517 | 12 | 23 |
| 8 | Mobile Networks and Applications | MONET | 7582 | 738 | 10.273 | 39 | 67 |
| 9 | Computer Networks | COMNET | 38758 | 3830 | 10.119 | 80 | 153 |
| 10 | Ad Hoc Networks | AHN | 3937 | 477 | 8.253 | 29 | 55 |
| 11 | IEEE Transactions on Mobile Computing | TMC | 5253 | 707 | 7.429 | 33 | 56 |
| 12 | Journal of High Speed Networks | JHSN | 1681 | 243 | 6.91 | 15 | 38 |
| 13 | IEEE Transactions on Communication | TCOM | 70982 | 10360 | 6.851 | 98 | 162 |
| 14 | ACM Transactions on Sensor Networks | MONET | 1074 | 174 | 6.712 | 17 | 27 |
| 15 | Queuing Systems- Theory and Applications | QUESTA | 6944 | 1179 | 5.889 | 33 | 54 |
| 16 | Networks | NETWORKS | 2890 | 745 | 3.879 | 22 | 30 |
| 17 | Journal of Network & Computer Applications | JNCA | 1645 | 439 | 3.747 | 15 | 33 |
| 18 | IEEE Transactions on Wireless Communications | TWC | 10086 | 2876 | 3.506 | 39 | 60 |
| 19 | Telecommunication Systems | TELESYS | 2565 | 765 | 3.352 | 24 | 37 |
| 20 | Computer Communications | COMCOM | 9497 | 3964 | 2.395 | 34 | 56 |

TABLE III
THE IMPACT FACTOR, $h$-INDEX, AND $g$-INDEX OF JOURNALS IN NETWORKS AND COMMUNICATION GROUP WITH CORRESPONDING ANALYTICAL DETAILS (JOURNALS ARE ARRANGED IN THE INCREASING NUMBER OF CITATIONS).

| Journal | $\sum_i c_i$ | $P$ | $I_f$ | $h$ | $g$ | $h^2$ | $g^2$ | $\sum_{i=g+1}^{P} c_i$ | $\sum_{i=1}^{h}(c_i-h)$ $+\sum_{i=h+1}^{P} c_i$ | $g^2 - h^2$ |
|---|---|---|---|---|---|---|---|---|---|---|
| JCN | 645 | 56 | 11.517 | 12 | 23 | 144 | 529 | 116 | 501 | 385 |
| TOSN | 1074 | 174 | 6.712 | 17 | 27 | 289 | 729 | 345 | 785 | 440 |
| JNCA | 1645 | 439 | 3.747 | 15 | 33 | 225 | 1089 | 556 | 1420 | 864 |
| JHSN | 1681 | 243 | 6.91 | 15 | 38 | 225 | 1444 | 237 | 1460 | 1219 |
| TELESYS | 2565 | 765 | 3.352 | 24 | 37 | 576 | 1369 | 1196 | 1989 | 793 |
| NETWORKS | 2890 | 745 | 3.879 | 22 | 30 | 484 | 900 | 1990 | 2406 | 416 |
| AHN | 3937 | 477 | 8.253 | 29 | 55 | 841 | 3025 | 912 | 2096 | 2184 |
| TMC | 5253 | 707 | 7.429 | 33 | 56 | 1089 | 3136 | 2117 | 4164 | 2047 |
| MCCR | 5304 | 434 | 12.221 | 35 | 66 | 1225 | 4356 | 948 | 4079 | 3131 |
| QUESTA | 6944 | 1179 | 5.889 | 33 | 54 | 1089 | 2916 | 4028 | 5855 | 1827 |
| MONET | 7582 | 738 | 10.273 | 39 | 67 | 1521 | 4489 | 3039 | 6061 | 2968 |
| COMCOM | 9497 | 3964 | 2.395 | 34 | 56 | 1156 | 3136 | 6361 | 8341 | 1980 |
| TWC | 10086 | 2876 | 3.506 | 39 | 60 | 1521 | 3600 | 6486 | 7210 | 2079 |
| WINET | 13588 | 894 | 15.199 | 49 | 102 | 2401 | 10404 | 3184 | 11187 | 8003 |
| PER | 27795 | 1630 | 17.052 | 76 | 132 | 5776 | 17424 | 10371 | 22019 | 11648 |
| COMNET | 38758 | 3830 | 10.119 | 80 | 153 | 6400 | 23409 | 15349 | 32358 | 17009 |
| ToN | 49148 | 1628 | 30.189 | 99 | 188 | 9801 | 35344 | 13804 | 39347 | 25543 |
| JSAC | 64330 | 3441 | 18.695 | 104 | 168 | 10816 | 28224 | 36106 | 53514 | 17408 |
| TCOM | 70982 | 10360 | 6.851 | 98 | 162 | 9604 | 26244 | 44738 | 61378 | 16640 |
| CCR | 85809 | 1813 | 47.33 | 150 | 203 | 22500 | 41209 | 44600 | 63309 | 18709 |

Figure 3 shows impact factor and indices as a function of the total number of citations of journals. Here, journals are arranged in the ascending order of their total number of citations, $\sum_i c_i$. We observe that $h$-index and $g$-index, generally, increase with an increase in the total number of citations. However, this is not true for the impact factor because it depends on the number of citations as well as the number of paper published. For journals with more number of papers published and whose number of citation are not so large, the impact factor is low. However, increasing the number of citations helps gaining some papers enough number of citations resulting an increase in the $h$-index as well as the $g$-index. As opposed to the indices, impact factor represents the quality of a journal in totality, therefore, it might not have increased in the same proportion as that of the indices.

Table III shows impact factor and indices of journals in the increasing order of their total number of citations. The values of $\sum_{i=g+1}^{P} c_i$, the difference of the summations of citations beyond the indices (as given by (19)), are given. Also, we listed the values for the difference of the squares of the indices in addition to the values of the parameters already listed in Table III.

Figure 4 shows the values of $h^2$, $g^2$, and $g^2 - h^2$ as a function of the total number of citations, where journals are arranged according to the increasing number of their citations. Also, it contains the number of papers published by the respective journal. We observe that there is a decrease in the values of $g^2 - h^2$ at some places. A closer look reveals that the decrease in the values of $g^2 - h^2$ is mainly due to the following reasons: (i) a decrease in the value of $g$-index for the respective journal, and/or (ii) an increase in the number of papers published by the respective journal, and the total number of citations for the respective journal might not have increased in the same proportion as that of the citations.

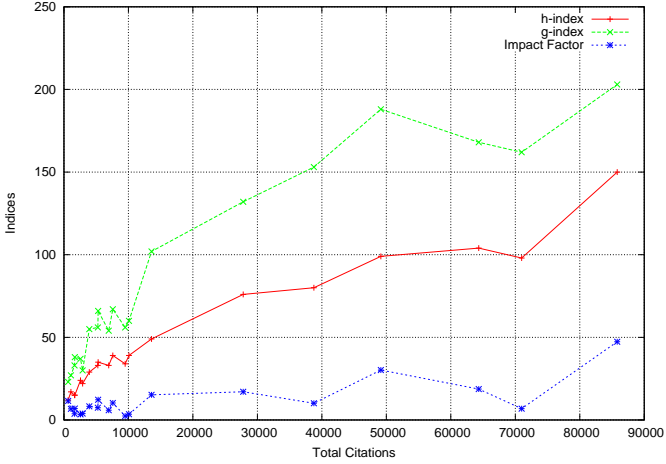

Fig. 3. The impact factor, $h$-index, and $g$-index as a function of the total number of citations, where journals are arranged according to increasing number of citations.

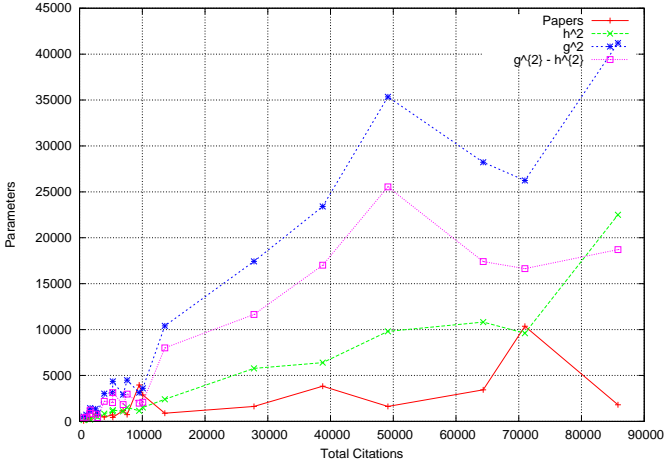

Fig. 4. The values of $h^2$, $g^2$, and $g^2 - h^2$ as a function of the total number of citations, where journals are arranged according to increasing number of citations.

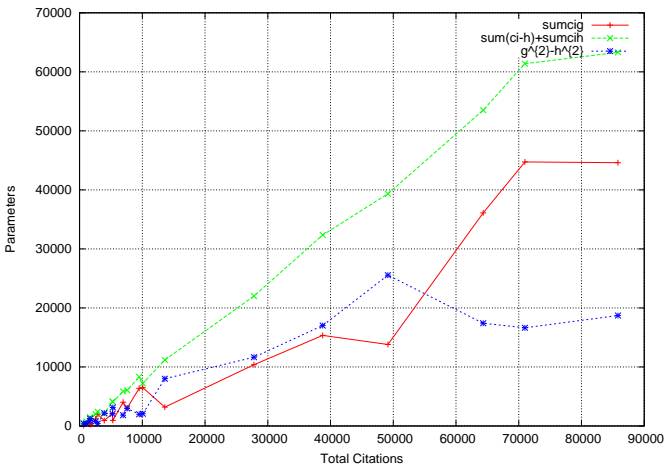

Fig. 5. The values of $g^2 - h^2$ and its constituents, namely, $\sum_{i=g+1}^{P} c_i$, $\sum_{i=1}^{P}(c_i - h) - \sum_{i=h+1}^{P} c_i$, as a function of the total number of citations, where journals are arranged according to increasing number of citations.

Figure 5 the values of $g^2 - h^2$ and its constituents, namely, $\sum_{i=g+1}^{P} c_i$, $\sum_{i=1}^{P}(c_i - h) + \sum_{i=h+1}^{P} c_i$, as a function of the total number of citations, where journals are arranged according to increasing number of their citations. We observe that the value of $g^2 - h^2$ for the sequence of journals, arranged in the increasing order of their citations, decreases for a journal if $\sum_{i=g+1}^{P} c_i$ decreased. Also, the value of $\sum_{i=1}^{P}(c_i - h) + \sum_{i=h+1}^{P} c_i$ increases with an increase in the total number of citations of the sequence of journals with a few exceptions. The reason for the decrease at some places forming an exception is an increase in the number of papers published by the journals appearing at those exceptional places, and the total number of citations have not increased in the same proportion.

## VI. CONCLUSION

In this paper, we presented an analysis of the relationships among the generalized impact factor, $h$-index, and $g$-index. Starting from the basic definitions of $h$-index, $g$-index, and the generalized impact factor, we derived mathematical equations relating these parameters. In an attempt to validate the relationships, we computed these parameters for *networks and communication* group in the area of computer science. We observed that journals which have a larger value of the generalized impact factor, also possess larger values of $h$-index and $g$-index (and vice versa), except in few cases. The exceptions are the journals with a large number of citations and a large number of papers published. These journals have enough number of highly cited papers to increase the $h$-index and $g$-index, even though they possess a relatively small impact factor. Another factor is that the number of citations might not have increased in the same proportions as that of the number of papers published by the journal. Further validations for different research domains form the future works.

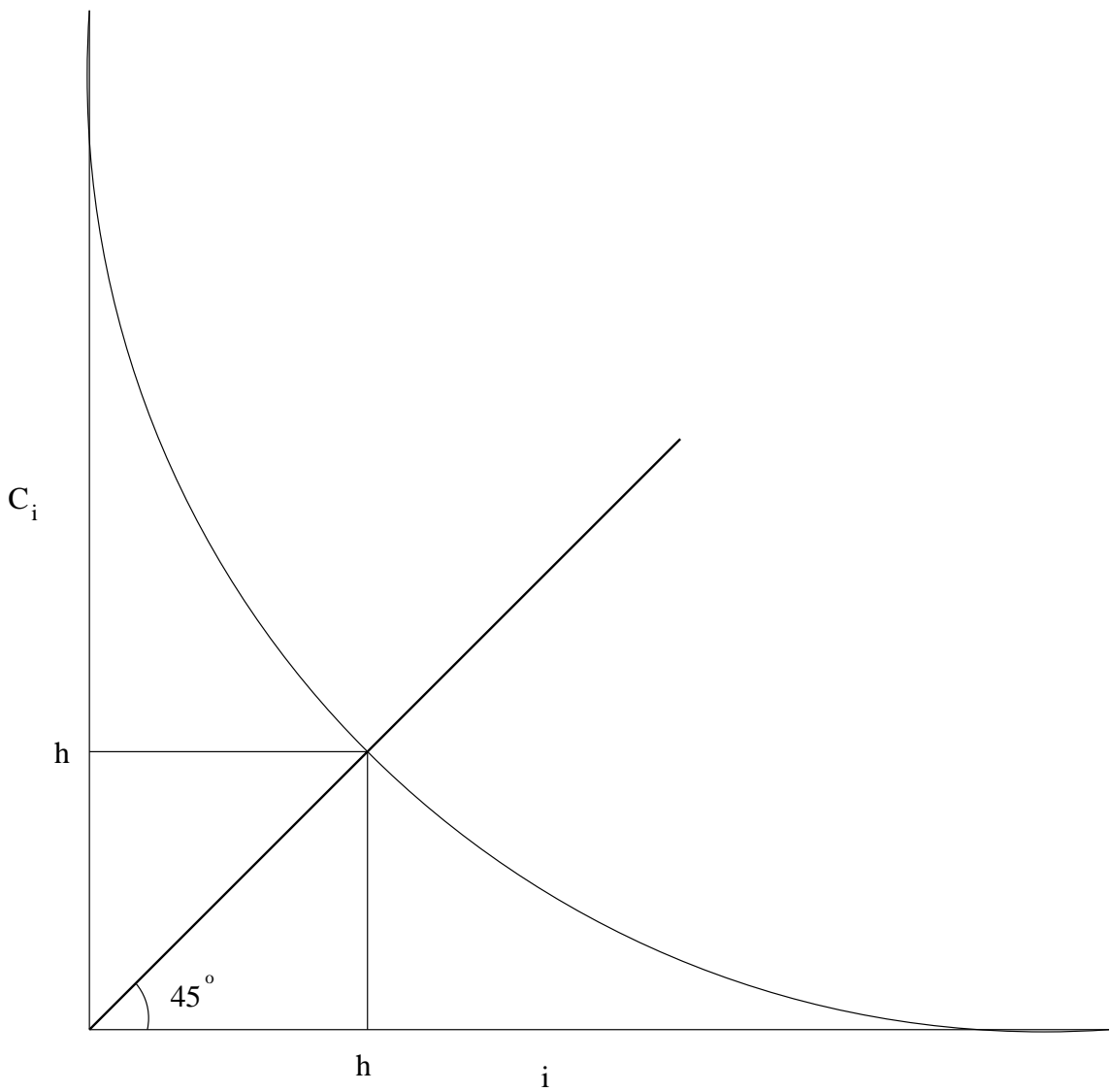